\documentclass[reprint,amssymb,amsmath,superscriptaddress,aps,showpacs,10pt,floatfix,longbibliography,prm]{revtex4-1}

\usepackage[pdftex]{graphicx}
\usepackage{epstopdf}
\usepackage{verbatim}
\usepackage{amsmath}
\usepackage{color}
\usepackage{subfigure}
\usepackage{amsbsy}
\usepackage{wasysym}
\usepackage{textcomp}
\usepackage{times}
\usepackage{float}
\usepackage{latexsym,amsmath,amssymb,bm,euscript}
\usepackage[colorlinks=true,linkcolor=black,citecolor=black,urlcolor=black]{hyperref}
\usepackage{hyperref}
\usepackage{soul}
\usepackage[normalem]{ulem}
\usepackage{mathrsfs}
\usepackage{lettrine}
\usepackage{xspace}
\usepackage{filecontents}

\begin{document}

\title{Proximate Tomonaga-Luttinger liquid in a spin-1/2 ferromagnetic XXZ chain compound}

\author{Boqiang Li}
\thanks{These authors contributed equally to this work}
\affiliation{Wuhan National High Magnetic Field Center and School of Physics, Huazhong University of Science and Technology, 430074 Wuhan, China}

\author{Xun Chen}
\thanks{These authors contributed equally to this work}
\affiliation{Wuhan National High Magnetic Field Center and School of Physics, Huazhong University of Science and Technology, 430074 Wuhan, China}

\author{Yuqian Zhao}
\thanks{These authors contributed equally to this work}
\affiliation{Wuhan National High Magnetic Field Center and School of Physics, Huazhong University of Science and Technology, 430074 Wuhan, China}

\author{Zhaohua Ma}
\affiliation{Wuhan National High Magnetic Field Center and School of Physics, Huazhong University of Science and Technology, 430074 Wuhan, China}

\author{Zongtang Wan}
\affiliation{Wuhan National High Magnetic Field Center and School of Physics, Huazhong University of Science and Technology, 430074 Wuhan, China}

\author{Yuesheng Li}
\email{yuesheng\_li@hust.edu.cn}
\affiliation{Wuhan National High Magnetic Field Center and School of Physics, Huazhong University of Science and Technology, 430074 Wuhan, China}

\date{\today}

\begin{abstract}
The spin-1/2 ferromagnetic XXZ chain is a prototypical many-body quantum model, exactly solvable via the integrable Bethe ansatz method, hosting a Tomonaga-Luttinger spin liquid. However, its clear experimental realizations remain absent. Here, we present a thorough investigation of the magnetism of the structurally disorder-free compound LuCu(OH)$_3$SO$_4$. By conducting magnetization and electron-spin-resonance measurements on the single-crystal sample, we establish that the title compound approximates the spin-1/2 ferromagnetic XXZ chain model with a nearest-neighbor exchange strength of $J_1$ $\sim$ 65 K and an easy-plane anisotropy of $\sim$ 0.994. The specific heat demonstrates a distinctive power-law behavior at low magnetic fields (with energy scales $\leq$ 0.02$J_1$) and low temperatures ($T$ $\leq$ 0.03$J_1$). This behavior is consistent with the expectations of the ideal spin-1/2 ferromagnetic XXZ chain model, thereby supporting the formation of a gapless Tomonaga-Luttinger spin liquid in LuCu(OH)$_3$SO$_4$.

\end{abstract}

\maketitle

\section{Introduction}

Spin-1/2 many-body correlated systems can give rise to exotic phases and elementary excitations due to their inherent strong quantum fluctuations and entanglements, capturing significant attention in condensed matter physics and materials science. The exploration, understanding, and control of these exotic phases have been a longstanding goal with promising applications in various fields, including future topological quantum computation~\cite{nayak2008non}. However, achieving this task is exceptionally challenging, stemming from two key aspects: 1. The majority of many-body-correlated quantum systems are too complex to have exact solutions, including eigenstates and eigenvalues. For instance, the ground-state nature of the spin-1/2 kagome-lattice Heisenberg antiferromagnet has been extensively studied for many years but remains elusive~\cite{Yan2011,PhysRevLett.109.067201,PhysRevLett.98.117205,PhysRevLett.118.137202}. 2. Real strongly correlated materials are so complex that their properties cannot be fully captured by simplified theoretical models, particularly in systems with low-lying small-gapped or gapless excitations. The Kitaev model is one of the few exactly solvable spin-1/2 many-body models~\cite{KITAEV20062}, but its experimental realizations are still under heated debate due to the widespread presence of significant non-Kitaev interactions in real materials~\cite{PhysRevB.96.064430,PhysRevLett.123.197201}.

The spin-1/2 XXZ or XXX (Heisenberg) chain, originally proposed by Werner Heisenberg and Hans Bethe, is a prototypical many-body model with eigenstates and eigenvalues that can be exactly solved using the integrable Bethe ansatz method~\cite{bethe1931theorie,heisenberg1985theorie}. In the case of a ferromagnetic nearest-neighbor (NN) exchange interaction (strength: $J_1$) and an easy-plane anisotropy (0 $<$ $\Delta$ $<$ 1), the model resides deeply within a gapless Tomonaga-Luttinger spin-liquid phase~\cite{PhysRevB.81.094430,PhysRevB.86.094417}. Numerous candidate compounds for the spin-1/2 ferromagnetic XXZ chain have been previously reported, such as LiCuVO$_4$~\cite{PhysRevLett.104.237207,PhysRevLett.106.219701,PhysRevLett.109.027203,nawa2013anisotropic,ruff2019chirality}, LiCuSbO$_4$~\cite{PhysRevLett.108.187206,grafe2017signatures,PhysRevB.96.224424}, Li$_2$ZrCuO$_4$~\cite{dussarrat2002synthesis,PhysRevLett.98.077202}, Li$_2$CuO$_2$~\cite{sapina1990crystal,lorenz2009highly}, Ca$_2$Y$_2$Cu$_5$O$_{10}$~\cite{PhysRevLett.109.117207,Thar2006NeutronSI,PhysRevB.100.104415}, CuAs$_2$O$_4$~\cite{PhysRevB.89.014412}, PbCuSO$_4$(OH)$_2$~\cite{PhysRevLett.108.117202,PhysRevB.106.144409}, and others. However, these candidates consistently show substantial next-nearest-neighbor (NNN) antiferromagnetic coupling ($J_2$), comparable to $J_1$. Despite their significance as strongly frustrated $J_1$-$J_2$ systems, the presence of substantial $J_2$ typically hinders exact resolution~\cite{WANG2022115663}. Furthermore, in the aforementioned Li-based compounds~\cite{nawa2013anisotropic,PhysRevLett.108.187206,dussarrat2002synthesis} and Ca$_2$Y$_2$Cu$_5$O$_{10}$ (Ca/Y disorder)~\cite{Thar2006NeutronSI}, structural disorders are inherent and unavoidable. These disorders can introduce significant complications for many-body modelling, given the apparent interaction randomness~\cite{PhysRevLett.118.107202,Li2019YbMgGaO4,PhysRevX.10.011007,Li2021spin,Lu2022}.

\begin{figure*}
\begin{center}
  \includegraphics[width=12cm,angle=0]{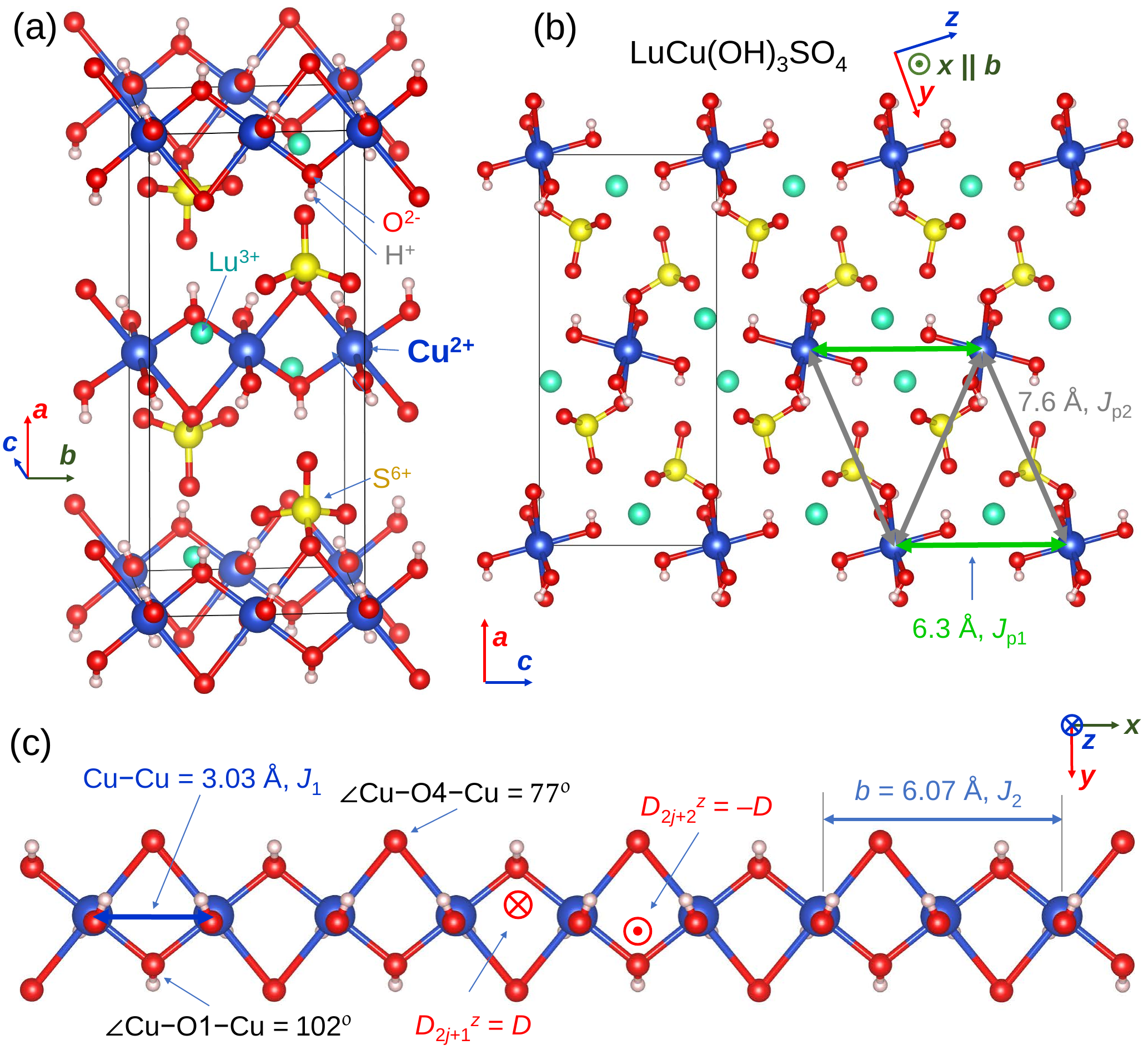}
  \caption{(a) Crystal structure of LuCu(OH)$_3$SO$_4$. (b) View along the spin chain ($b$ axis) of the crystal structure. (c) CuO$_2$ chain viewed along the $z$ axis. Dominant nearest-neighbor (NN) ferromagnetic exchange ($J_1$) and weaker couplings beyond first neighbors ($J_2$, $J_\mathrm{p1}$, and $J_\mathrm{p2}$) are displayed in (b) and (c). NN Dzyaloshinsky-Moriya interaction vectors are shown in (c). The local $xyz$-coordinate system for spin components is defined, and thin black lines mark the unit cells.}
  \label{fig1}
\end{center}
\end{figure*}

Recently, Huangjie Lu et al. reported the successful synthesis of the structurally disorder-free Cu$^{2+}$ (spin-1/2) chain compound LuCu(OH)$_3$SO$_4$, featuring a crystal size of approximately 10$\times$25$\times$10 $\mu$m~\cite{lu2020structural}. However, to the best of our knowledge, its magnetism has not been reported so far. In this study, we implemented a new synthesis method and achieved successful growth of significantly larger single crystals of LuCu(OH)$_3$SO$_4$, typically with a size of $\sim$ 0.08$\times$1$\times$0.08 mm. Through magnetization and electron-spin-resonance (ESR) measurements on the single-crystal sample, we determined the detailed spin Hamiltonian. Our findings indicate that both interchain and next-nearest-neighbor intrachain couplings are less significant, $J_\mathrm{p1}$/$J_1$ $\sim$ 1.01\%, $J_\mathrm{p2}$/$J_1$ $\sim$ $-$0.18\%, and $J_2$/$J_1$ $\sim$ 1\% (see Fig.~\ref{fig1}). LuCu(OH)$_3$SO$_4$ closely approximates the ideal spin-1/2 ferromagnetic XXZ chain model with $J_1$ $\sim$ 65 K and $\Delta$ $\sim$ 0.994. The specific heat and magnetization measurements down to $T$ = 35 mK ($\sim$ 0.0005$J_1$) are consistent with the expectations of the ideal model, supporting the emergence of a gapless Tomonaga-Luttinger spin liquid in LuCu(OH)$_3$SO$_4$.

\begin{figure*}
\begin{center}
  \includegraphics[width=14cm,angle=0]{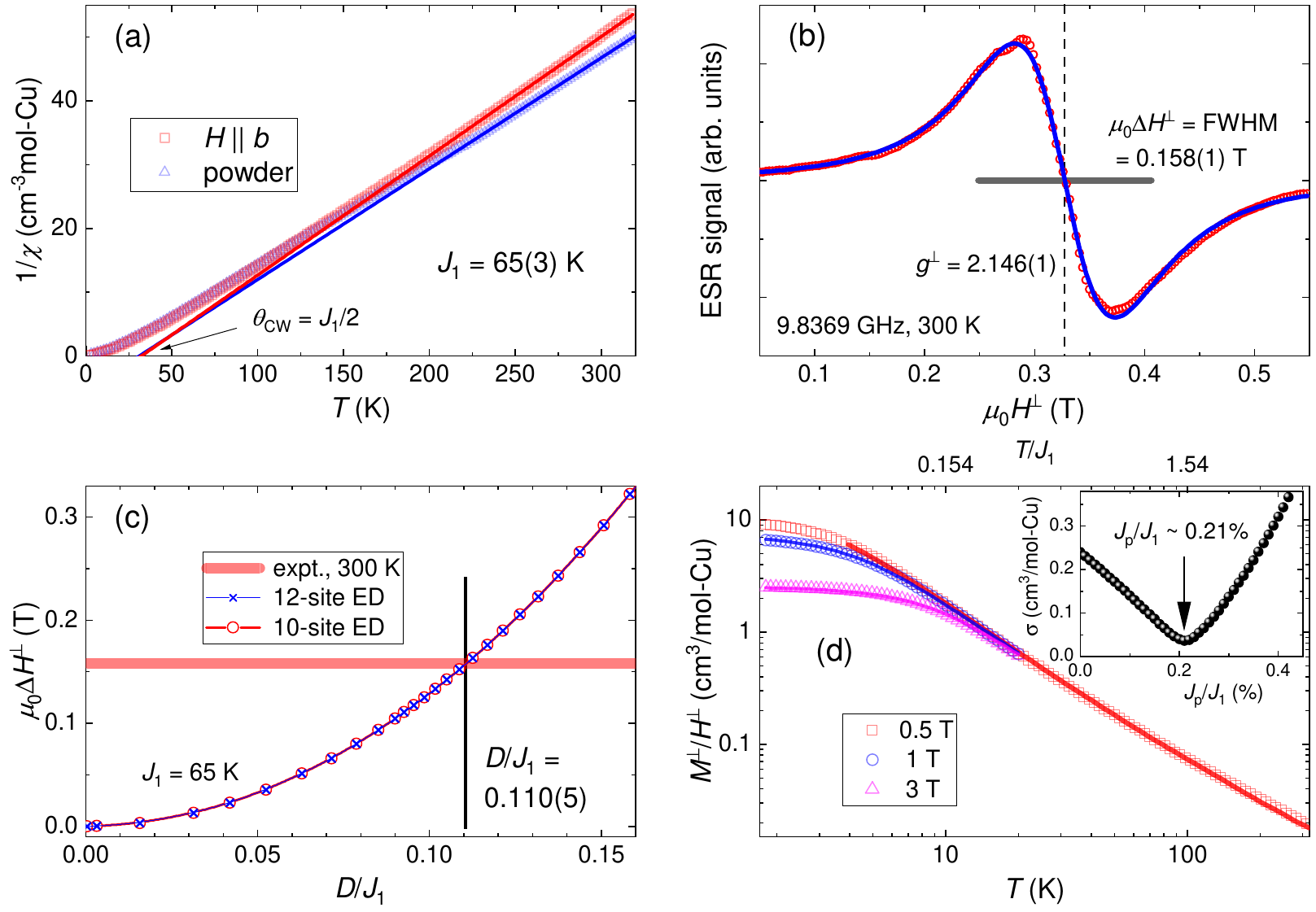}
  \caption{Determining couplings $J_1$, $D$, and $J_\mathrm{p}$ $\equiv$ $J_\mathrm{p1}$/3+2$J_\mathrm{p2}$/3. (a) Inverse magnetic susceptibilities measured on single-crystal and powder samples, with Curie-Weiss fits (lines) above 200 K. (b) Electron spin resonance (ESR) spectrum measured along the $b$ axis ($H^\perp$) on the single-crystal sample at 300 K. The blue line represents the fit using the derivative Lorentzian function, with resonance field ($g^\perp$ = 2.146) and full width at half maximum (FWHM = $\mu_0\Delta H^\perp$) marked. (c) Dzyaloshinsky-Moriya (DM) strength ($D$) dependence of $\mu_0\Delta H^\perp$, calculated using exact diagonalization (ED) with $J_1$ = 65 K and at $T$ = 300 K. The determined DM anisotropic strength is $D$/$J_1$ = 0.110(5). (d) Magnetization measured on the single-crystal sample with magnetic fields along the $b$ axis, with stochastic series expansion (SSE) calculations (lines) shown using least-square parameters. Inset: Standard deviation $\sigma$ as a function of $J_\mathrm{p}$/$J_1$ calculated using the experimental data between 1.8 and 20 K at $\mu_0H^\perp$ = 1 and 3 T. The arrow denotes the optimal parameter $J_\mathrm{p}$/$J_1$ = 0.21(1)\%. Periodic boundary conditions are used in both ED and SSE computations.}
  \label{fig2}
\end{center}
\end{figure*}

\section{METHODS}

\subsection{Sample synthesis}

The microcrystals of LuCu(OH)$_3$SO$_4$ were synthesized using the traditional hydrothermal method as reported in Ref.~\cite{lu2020structural}, resulting in crystals of approximately 5$\times$30$\times$5 $\mu$m in size. To enhance crystal size, we employed copper plates (Cu $\geq$ 99.99\%) as the Cu source to slow down the chemical reaction. This modification successfully increased the crystal size to $\sim$ 0.08$\times$1$\times$0.08 mm~\cite{supple}. The high quality and phase purity of our samples were confirmed through both powder and single-crystal XRD measurements. The longest dimension of the crystal aligns with the $b$ axis, and approximately 800 of the largest single crystals, totaling $\sim$ 10 mg, are aligned along the $b$ axis using GE varnish for magnetic property measurements. We are only able to align the crystals along the $b$ axis due to the small dimensions that make distinguishing non-$b$ directions challenging.

\subsection{Sample characterizations}

The ESR spectra were obtained using a continuous-wave spectrometer (Bruker EMXmicro-6/1) at x-band frequencies ($\sim$ 9.8 GHz) for both single-crystal and powder samples ($\sim$ 10 mg). Magnetization and dc/ac susceptibilities of LuCu(OH)$_3$SO$_4$ (up to 7 T, down to 1.8 K) were measured in a magnetic property measurement system (Quantum Design) using powder samples ($\sim$ 38 mg, ground from single crystals) and well-aligned single-crystal samples ($\sim$ 10 mg) along the $b$ axis. The powder sample (6.95 mg) was utilized for magnetization measurements at between 0.035 and 1.8 K, employing a Faraday force magnetometer within a $^3$He-$^4$He dilution refrigerator (KELMX-400, Oxford Instruments). Superconducting coils (in INTA-LLD-S12/14, Oxford Instruments) generated the main magnetic fields and field gradients (d$H$/d$l$ = $\pm$ 10 T/m), and electrical capacitance was measured using a digital capacitance bridge (AH-2500A, Andeen-Hagerling, Inc.) with the three-terminal method~\cite{shimizu2021development,PhysRevLett.122.137201}.

The specific heat measurements, down to 1.8 K, were performed in a Physical Properties Measurement System (Quantum Design) using dry-pressed disks of the powders (4.86 mg). N grease was used to facilitate thermal contact between the power sample and the puck, and the sample coupling was measured to be better than 95\%. The specific-heat contributions of the N grease and puck were initially measured and subtracted from the data. Specific heat measurements between 0.1 and 1.8 K were carried out using a thermal relaxation method in the dilution refrigerator, utilizing a powder sample weighing 2.61 mg. The thermal link to the bath was established using a bronze wire of appropriate diameter and length to achieve quasiadiabatic conditions~\cite{PhysRevB.97.184434,PhysRevX.10.011007}. A $\sim$ 10 nm NiCr layer was deposited on one side of the platform and served as a heater (resistance $\sim$ 10 k$\Omega$). The chip thermometer (CX-1010-BR, LakeShore) was \emph{in situ} calibrated against a reference thermometer (RX-102B-RS-0.02B, LakeShore, calibrated down to 20 mK) at each applied magnetic field by turning off the heater.

\subsection{Numerical methods}

We utilized finite-size ($N \leq 28$) ED, large-scale ($N = 100$) stochastic series expansion (SSE), and integrable Bethe ansatz ($N \to \infty$) methods to thoroughly investigate the strongly correlated quantum magnetism of LuCu(OH)$_3$SO$_4$. In the Bethe simulations, we employed $\Delta = \cos(\pi/29)$ ($\sim$ 0.994). Detailed discussions on the validation, consistency, and finite-size effects of these three methods are provided in Supplementary material~\cite{supple}.

\section{Results and Discussions}

\subsection{Crystal structure and symmetry analysis}

Figure~\ref{fig1} displays the crystal structure of LuCu(OH)$_3$SO$_4$ refined using single-crystal x-ray diffraction (XRD) data~\cite{supple}. Due to significant differences among Cu$^{2+}$, Lu$^{3+}$, OH$^{-}$, and SO$_4^{2-}$, the occurrence of site-mixing structural disorder is expected to be prohibited. Firstly, our XRD data, measured on both powder and single-crystal samples~\cite{supple}, indeed show no clear evidence of structural disorder, aligning with previously reported results~\cite{lu2020structural}. Secondly, the single-crystal ESR spectrum, highly sensitive to quasi-free magnetic moments induced by structural disorder and interaction randomness~\cite{arXiv:2107.12712}, does not exhibit any sharp components (linewidth $<$ 0.02 T) with hyperfine structures [see Fig.~\ref{fig2}(b)]. Lastly, the single-crystal magnetization can be well reproduced by the \emph{ab-initio} model down to $\sim$ 1.8 K ($\mu_0H^\perp$ $\geq$ 1 T) without considering structural disorder [see Fig.~\ref{fig2}(d)]. Therefore, we propose LuCu(OH)$_3$SO$_4$ as a structurally disorder-free compound for the study of strongly correlated quantum magnetism.

The superexchange coupling ($J_1$) between the NN Cu$^{2+}$ ions, with Cu-Cu distance $b$/2 = 3.03 \AA, is mediated by O1$^{2-}$ and O4$^{2-}$ along the spin-1/2 chain, with $|$Cu-O1$|$ $\sim$ 1.95 \AA~\& $\angle$Cu-O1-Cu $\sim$ 102$^\circ$ and $|$Cu-O4$|$ $\sim$ 2.44 \AA~\& $\angle$Cu-O4-Cu $\sim$ 77$^\circ$ (see Fig.~\ref{fig1}). When the Cu-O-Cu bond angle exceeds the critical value of $\varphi_\mathrm{c}$ $\sim$ 96$^\circ$, the exchange interaction tends to be antiferromagnetic (AF). Conversely, a ferromagnetic (FM) exchange interaction is expected when the bond angle is smaller. In LuCu(OH)$_3$SO$_4$, the superexchanges along the Cu-O1-Cu (AF) and Cu-O4-Cu (FM) paths may compete. Given that $\varphi_\mathrm{c}-\angle$Cu-O4-Cu $\sim$ 19$^\circ$ exceeds $\angle$Cu-O1-Cu$-\varphi_\mathrm{c}$ $\sim$ 6$^\circ$, we find the detection of a ferromagnetic nature in $J_1$ [Fig.~\ref{fig2}(a)] within LuCu(OH)$_3$SO$_4$ to be unsurprising.

Magnetic chains are well spatially separated by nonmagnetic Lu$^{3+}$ and SO$_4^{2-}$ with much larger distances of 6.3 \AA ~along the $J_\mathrm{p1}$ path and 7.6 \AA ~along the $J_\mathrm{p2}$ path [see Fig.~\ref{fig1}(b)], respectively. This suggests the quasi-one-dimensional nature of the spin system in LuCu(OH)$_3$SO$_4$. Our density functional theory (DFT)+$U$ calculations for LuCu(OH)$_3$SO$_4$ semi-quantitatively demonstrate that $J_1$ $\sim$ 50 K is ferromagnetic, with $J_\mathrm{p1}$/$J_1$ $\sim$ 4.8\%, $J_\mathrm{p2}$/$J_1$ $\sim$ 0.1\%, and $J_2$/$J_1$ $\sim$ 1.0\%~\cite{supple}. These results indicate that the spin system of LuCu(OH)$_3$SO$_4$ closely approximates the ideal spin-1/2 ferromagnetic chain model.

In Cu$^{2+}$-based magnets, the NN exchange coupling is typically isotropic and symmetric (Heisenberg). In addition to the isotropic NN coupling, the antisymmetric Dzyaloshinsky-Moriya (DM) interaction tends to be dominant, arising from spin-orbit coupling if allowed by symmetry~\cite{PhysRev.120.91,PhysRevLett.101.026405}. Similar to other Cu$^{2+}$-based magnets, such as ZnCu$_3$(OH)$_6$Cl$_2$~\cite{PhysRevLett.101.026405} and YCu$_3$(OH)$_{6.5}$Br$_{2.5}$~\cite{arXiv:2107.12712}, we find, through a standard symmetry analysis, that the nearly Heisenberg Hamiltonian invariant under the $Pnma$ space group symmetry of LuCu(OH)$_3$SO$_4$ [see Fig.~\ref{fig1}(c)] is given by,
\begin{multline}
\mathcal{H} = -J_1\sum_j\mathbf{S}_j\cdot\mathbf{S}_{j+1}+D\sum_j(-1)^{j+1}(S_j^xS_{j+1}^y-S_j^yS_{j+1}^x)\\
+\mathcal{H}',
\label{eq1}
\end{multline}
where $\mathcal{H}'$ = $J_2\sum_j\mathbf{S}_j\cdot\mathbf{S}_{j+2}+J_\mathrm{p1}\sum_{\langle j,j\mathrm{p1}\rangle}\mathbf{S}_j\cdot\mathbf{S}_{j\mathrm{p1}}+J_\mathrm{p2}\sum_{\langle j,j\mathrm{p2}\rangle}\mathbf{S}_j\cdot\mathbf{S}_{j\mathrm{p2}}+...$ represents interactions beyond the nearest neighbors. The periodic boundary condition, $\mathbf{S}_j$ $\equiv$ $\mathbf{S}_{j+N}$, is considered, where $N$ is an even~\footnote{In LuCu(OH)$_3$SO$_4$, each unit cell contains two spins along each chain.} number of spins (i.e., spin length). Since inversion centers are located at any Cu$^{2+}$ ions, the NNN DM interaction is symmetrically forbidden. The parameter $D$ in Equation (\ref{eq1}) denotes the strength of the NN DM interaction, and its sign does not affect the resulting observables due to symmetry. Therefore, for consistency, we set $D$ $>$ 0 throughout this work. Additionally, the superscripts, ``$\perp$" and ``$||$", denote the quantity with the applied magnetic field perpendicular and parallel to the $z$ axis of the spin model [see Fig.~\ref{fig1}(c)], respectively, throughout this work.

Above $\sim$ 200 K, the inverse magnetic susceptibilities ($\chi$) of LuCu(OH)$_3$SO$_4$ exhibit a linear $T$-dependence, following the Curie-Weiss (CW) law, $\chi$ = $C$/($T-\theta_\mathrm{CW}$) [see Fig.~\ref{fig2}(a)]. Through CW fits, we obtain the $g$ factor as $g$ = $\sqrt{4k_\mathrm{B}C/(\mu_0N_\mathrm{A}\mu_\mathrm{B}^2)}$. The resulting values are $g^\perp$ $\sim$ 2.13 and $g^\mathrm{powder}$ $\sim$ 2.2, based on magnetic susceptibilities measured on the single-crystal and powder samples, respectively. These values are roughly consistent with the ESR [see Fig.~\ref{fig2}(b)] and magnetization data. Additionally, the NN ferromagnetic couplings are fitted to be $J_1$ = 2$\theta_\mathrm{CW}$ = 64.9$\pm$0.7 and 62.3$\pm$0.6 K for the single-crystal and powder samples, respectively [Fig.~\ref{fig2}(a)]. Thus, we conclude the ferromagnetic $J_1$ = 65(3) K for LuCu(OH)$_3$SO$_4$.

\begin{figure*}
\begin{center}
  \includegraphics[width=13cm,angle=0]{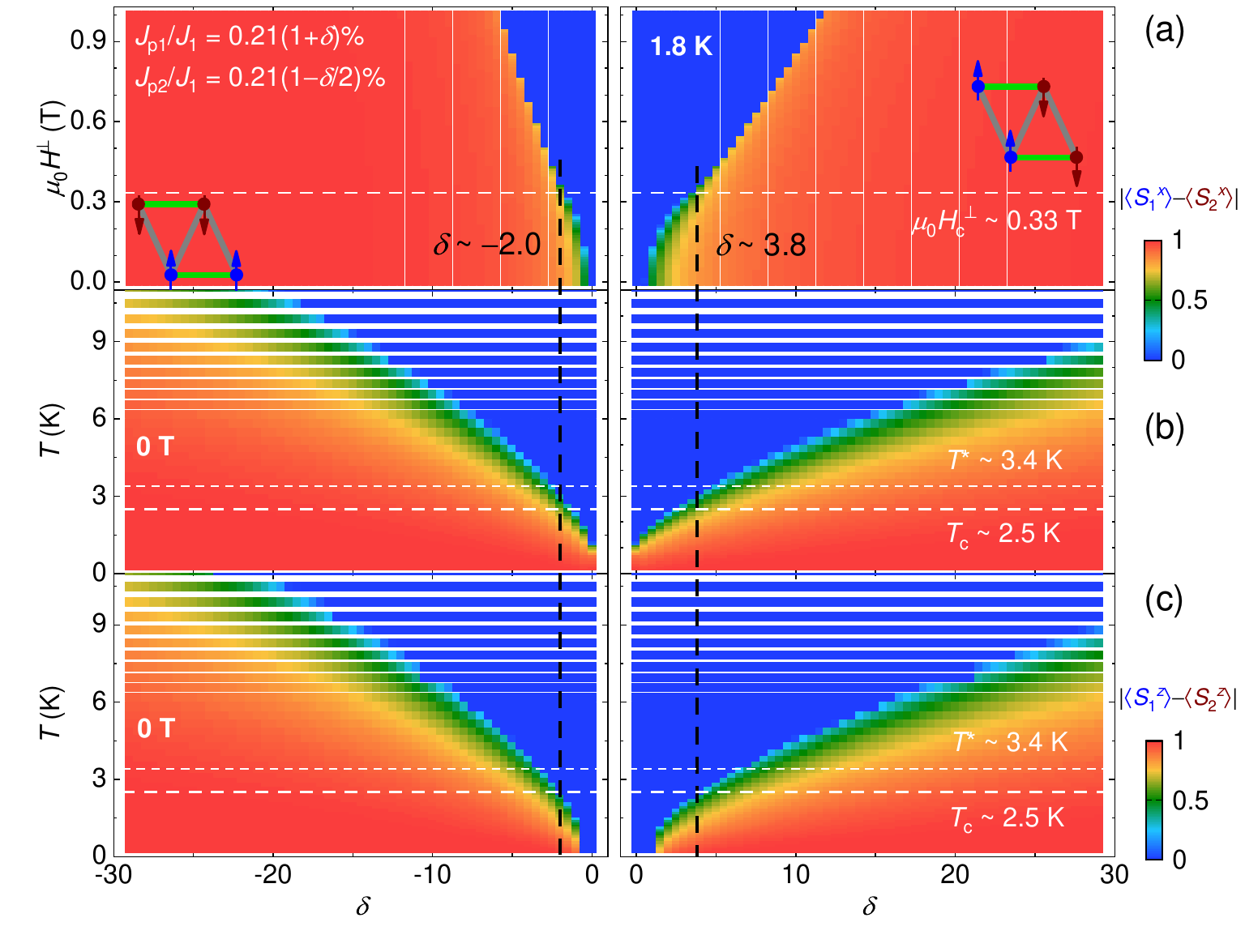}
  \caption{Interchain antiferromagnetic order parameters calculated using the SSE method with a self-consistent mean-field treatment of interchain interactions. (a) $|\langle S_1^x\rangle-\langle S_2^x\rangle|$ at 1.8 K. (b) $|\langle S_1^x\rangle-\langle S_2^x\rangle|$ at 0 T. (c) $|\langle S_1^z\rangle-\langle S_2^z\rangle|$ at 0 T. The normalized interchain couplings are given as $J_\mathrm{p1}$/$J_1$ = (1$+\delta$)$J_\mathrm{p}$/$J_1$ and $J_\mathrm{p2}$/$J_1$ = (1$-\delta$/2)$J_\mathrm{p}$/$J_1$, where $J_\mathrm{p}$/$J_1$ = 0.21\% (antiferromagnetic) is fixed (refer to Fig.~\ref{fig2}). At $\delta$ $>$ 0, the dominant antiferromagnetic $J_\mathrm{p1}$ induces interchain stripe-\uppercase\expandafter{\romannumeral1} order along $J_\mathrm{p1}$ (along green bonds, see right inset of a), whereas at $\delta$ $<$ 0, the dominant antiferromagnetic $J_\mathrm{p2}$ causes stripe-\uppercase\expandafter{\romannumeral2} order along $J_\mathrm{p2}$ (along grey bonds, see left inset of a). The critical field of $\mu_0H_\mathrm{c}^\perp$ $\sim$ 0.33 T (dashed white line in a) and critical temperature of $T_\mathrm{c}$ $\sim$ 2.5 K are obtained from the field and temperature dependence of susceptibility measured at $T$ = 1.8 K and $H^\perp$ $\rightarrow$ 0, respectively, on the single-crystal sample (refer to Fig.~\ref{fig4}). $T^*$ denotes the temperature at which interchain magnetic ordering starts [Fig.~\ref{fig4}(d)].}
  \label{fig3}
\end{center}
\end{figure*}

\begin{figure*}
\begin{center}
  \includegraphics[width=14cm,angle=0]{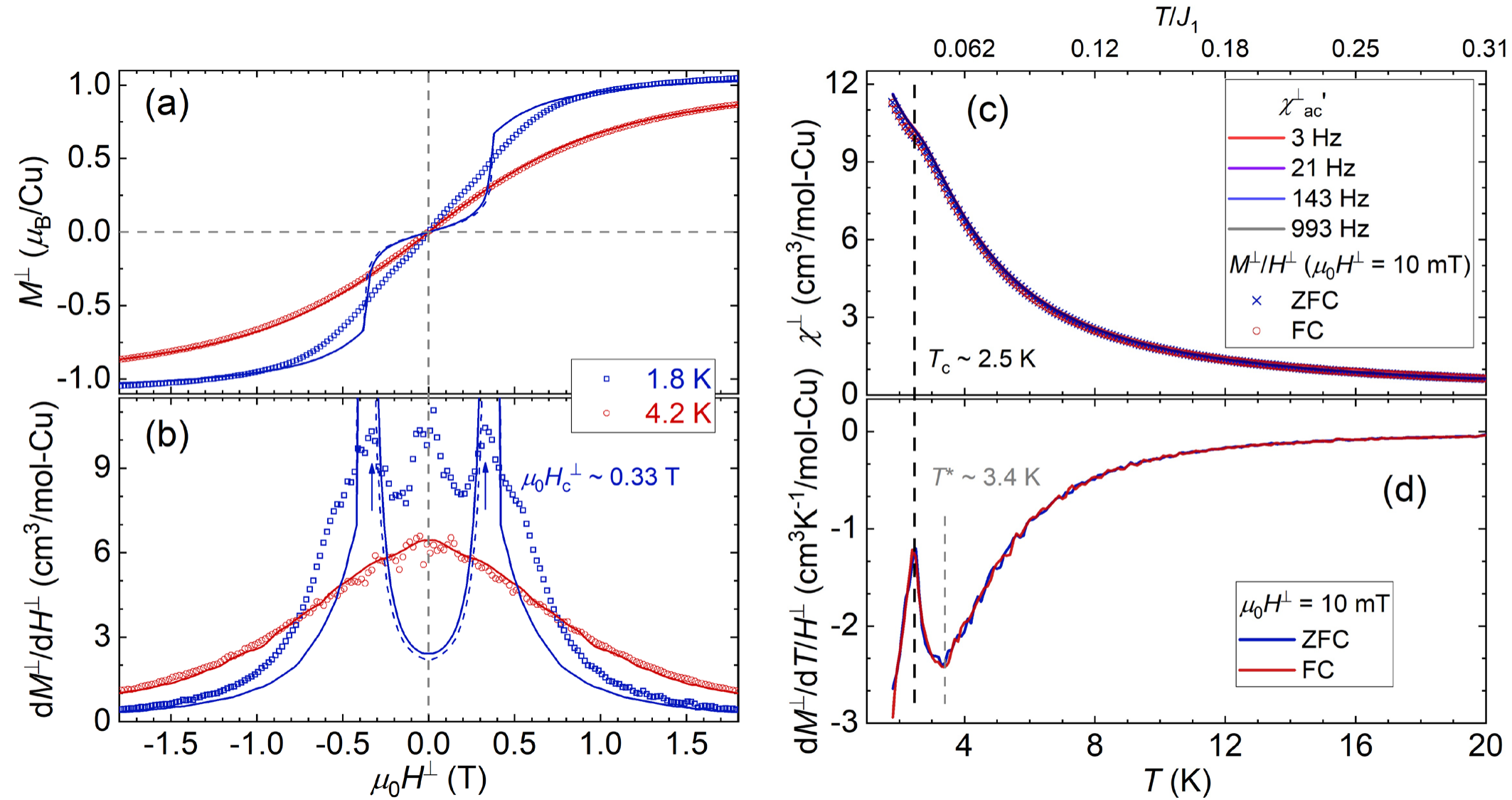}
  \caption{Magnetization measured on the single-crystal sample. Isothermal (a) magnetization ($M^\perp$) and (b) susceptibilities (d$M^\perp$/d$H^\perp$) along the $b$ axis. Solid and dashed colored lines represent SSE simulations with the self-consistent mean-field treatment of interchain interactions, using $\delta$ = 3.8 and $-$2.0, respectively; the solid and dashed red lines (at 4.2 K) overlap. The critical field of $\mu_0H_\mathrm{c}^\perp$ $\sim$ 0.33 T is indicated by arrows in (b). Although complete loops were measured, no evident magnetic hysteresis was detected. (c) Ac susceptibilities measured at various frequencies, and dc susceptibilities measured under zero-field cooling (ZFC) and field cooling (FC). (d) The $T$-derivative magnetization at $\mu_0H^\perp$ = 10 mT. The critical temperature $T_\mathrm{c}$ ($\sim$ 2.5 K) and the onset temperature $T^*$ ($\sim$ 3.4 K) are indicated by dashed lines.}
  \label{fig4}
\end{center}
\end{figure*}

\subsection{Electron spin resonance spectrum}

Figure~\ref{fig2}(b) displays the x-band ESR spectrum measured on the single-crystal sample of LuCu(OH)$_3$SO$_4$ with the magnetic field applied along the $b$ axis. The experimental data are well-fitted using the first derivative Lorentzian function [see Fig.~\ref{fig2}(b)]~\cite{PhysRevLett.115.167203}, revealing a sizable linewidth of $\mu_0\Delta H^\perp$ = 0.158(1) T. Below, we discuss various sources contributing to the broadening of the ESR signal in the high-$T$ limit.

Firstly, the hyperfine interaction contributes to the ESR linewidth as $\mu_0\Delta H_\mathrm{h}$ $\sim$ $|A_\mathrm{h}|^2$/($g\mu_\mathrm{B}J_1$) $<$ 0.01 mT, where $|A_\mathrm{h}|$ $<$ 500 MHz is the hyperfine coupling between Cu$^{2+}$ electronic and nuclear spins~\cite{Saladino2003}. Secondly, the dipole-dipole magnetic interaction also broadens the signal by $\mu_0\Delta H_\mathrm{d}$ $\sim$ $|E_\mathrm{d}|^2$/($g\mu_\mathrm{B}J_1$) $\sim$ 0.09 mT. Here, $|E_\mathrm{d}|$ $\sim$ 2$\mu_0g^2\mu_\mathrm{B}^2$/($\pi b^3$) $\sim$ 1.9 GHz represents the strength of the dipole-dipole interaction, and $b$/2 is the NN Cu-Cu distance [see Fig.~\ref{fig1}(c)]. Thirdly, as there is only one Wyckoff position for Cu$^{2+}$ ions, the uniform Zeeman interaction cannot broaden the single-crystal ESR signal~\cite{PhysRevLett.101.026405,PhysRevLett.115.167203,Liu2021Frustrated}. Even when all the above contributions are considered together, they are more than two orders of magnitude smaller than the observed linewidth. Therefore, in the high-$T$ limit ($T$ $\sim$ 300 K), the exchange anisotropy of the DM interaction predominantly accounts for the observed ESR linewidth in LuCu(OH)$_3$SO$_4$, with $\mu_0\Delta H^\perp$ $\sim$ $D^2$/($g\mu_\mathrm{B}J_1$) and $D$/$J_1$ $\sim$ 0.1, similar to several other Cu$^{2+}$-based magnets~\cite{PhysRevLett.101.026405}.

To determine $D$ precisely, we conduct detailed many-body calculations for the transverse ESR linewidth measured at 300 K using $\mu_0\Delta H^\perp$ = $\sqrt{2\pi(M_2^x)^3/M_4^x}/(g^\perp\mu_B)$. Here, $M_2^x$ = $\langle[\mathcal{H}_\mathrm{DM}^x,M^+][M^-,\mathcal{H}_\mathrm{DM}^x]\rangle$/$\langle M^+M^-\rangle$ and $M_4^x$ = $\langle[\mathcal{H}^x,[\mathcal{H}_\mathrm{DM}^x,M^+]][\mathcal{H}^x,[\mathcal{H}_\mathrm{DM}^x,M^-]]\rangle$/$\langle M^+M^-\rangle$ are the second and fourth moments, respectively~\cite{PhysRevB.4.38}. In these expressions, we set $\mathcal{H}_\mathrm{DM}^x$ = $D\sum_j(-1)^{j+1}(S_j^yS_{j+1}^z-S_j^zS_{j+1}^y)$, $\mathcal{H}^x$ = $-J_1\sum_j\mathbf{S}_j\cdot\mathbf{S}_{j+1}+\mathcal{H}_\mathrm{DM}^x-\mu_0H^\perp g^\perp\sum_jS_j^z$, $M^\pm$ $\equiv$ $\sum_jS_j^x\pm iS_j^y$, and $\langle\rangle$ represents the equilibrium thermal average, $\langle O\rangle$ = $\mathrm{Tr}(O\mathrm{e}^{-\beta\mathcal{H}^x})$/$\mathrm{Tr}(\mathrm{e}^{-\beta\mathcal{H}^x})$ where $\beta\equiv1/(k_\mathrm{B}T)$. At a high temperature of $T$ = 300 K ($\sim$ 4.6$J_1$), the calculated $\mu_0\Delta H^\perp$ shows no evident finite-size effect and is a monotone increasing function of $D$/$J_1$, as shown in Fig.~\ref{fig2}(c). By comparing with the experimental ESR linewidth, we conclude $D$/$J_1$ = 0.110(5) for LuCu(OH)$_3$SO$_4$. Furthermore, the observed $g$-factor shift from the free electron value $g_e$ = 2.0023, $(g^\perp-g_e)$/$g_e$ $\sim$ 0.072 ($\sim$ $D$/$J_1$)~\cite{PhysRev.120.91}, also supports the above conclusion.

In comparison, the main ESR signal of the powder sample is broader, $\mu_0\Delta H_\mathrm{powder}$ $\sim$ 0.36 T~\cite{supple}. The ESR linewidth calculated along the $z$ axis is two times larger than the transverse one, $\mu_0\Delta H^\parallel$ $\sim$ 2$\mu_0\Delta H^\perp$, which partially accounts for the larger value of $\mu_0\Delta H_\mathrm{powder}$. On the other hand, the anisotropy of the $g$-factor tensor should further explain the observed $\mu_0\Delta H_\mathrm{powder}$ in LuCu(OH)$_3$SO$_4$.

\subsection{Ferromagnetic XXZ chain model}

The original spin Hamiltonian of LuCu(OH)$_3$SO$_4$ [Eq.~(\ref{eq1})] includes the NN antisymmetric DM interaction, significantly increasing the computational cost and making low-$T$ simulations challenging. Therefore, we introduce a unitary transformation operator,
\begin{equation}
\mathcal{T}_\mathrm{u} = \exp(i\sum_j\phi_jS_j^z),
\label{eq2}
\end{equation}
where $\phi_j$ = arctan($D$/$J_1$)[1$-$($-$1)$^{j+1}$]/2. Through a canonical transformation, $\widetilde{\mathcal{H}}$ = $\mathcal{T}_\mathrm{u}^\dagger\mathcal{H}\mathcal{T}_\mathrm{u}$~\cite{PhysRevB.62.R751}, we obtain a spin-1/2 ferromagnetic XXZ chain,
\begin{multline}
\widetilde{\mathcal{H}} = -\sqrt{J_1^2+D^2}\sum_j(S_j^xS_{j+1}^x+S_j^yS_{j+1}^y+\Delta S_j^zS_{j+1}^z)
\\+\mathcal{H}',
\label{eq3}
\end{multline}
where $\Delta$ = $J_1$/$\sqrt{J_1^2+D^2}$ = 0.994 shows an easy-plane anisotropy for LuCu(OH)$_3$SO$_4$. Since the unitary transformation doesn't change the trace, one can calculate the observable by $\langle O\rangle$ = $\mathrm{Tr}(O\mathrm{e}^{-\beta\mathcal{H}})$/$\mathrm{Tr}(\mathrm{e}^{-\beta\mathcal{H}})$ = $\mathrm{Tr}(\widetilde{O}\mathrm{e}^{-\beta\widetilde{\mathcal{H}}})$/$\mathrm{Tr}(\mathrm{e}^{-\beta\widetilde{\mathcal{H}}})$, where $\widetilde{O}$ = $\mathcal{T}_\mathrm{u}^\dagger O\mathcal{T}_\mathrm{u}$. We confirmed the equivalence between Equations (\ref{eq1}) and (\ref{eq3}) through finite-size exact diagonalization (ED) calculations.

When a magnetic field is applied along the $z$ axis, the Zeeman term remains invariant under $\mathcal{T}_\mathrm{u}$, $\widetilde{\mathcal{H}_\mathrm{Z}^\parallel}$ $\equiv$ $\mathcal{H}_\mathrm{Z}^\parallel$, and thus we can efficiently utilize the ferromagnetic XXZ chain model expressed in Eq. (\ref{eq3}) along with $\mathcal{H}_\mathrm{Z}^\parallel$ to calculate observables. In cases where the magnetic field is applied perpendicular to the $z$ axis, such as along the $b$ axis, the Zeeman term is no longer invariant under $\mathcal{T}_\mathrm{u}$,  leading to expensive computations. Fortunately, for LuCu(OH)$_3$SO$_4$ with a small tilt angle $\phi_0$ = arctan($D$/$J_1$) $\sim$ 6$^\circ$, the differences in observables using $\widetilde{\mathcal{H}}$ (Equation (\ref{eq3})) plus $\widetilde{\mathcal{H}_\mathrm{Z}^\perp}$ (DM, exact) and $\mathcal{H}_\mathrm{Z}^\perp$ (XXZ, effective) are negligible within the temperature-field range of our experiments. For magnetization/susceptibility, we find 1 $<$ $M_\mathrm{XXZ}^\perp$/$M_\mathrm{DM}^\perp$ $\leq$ 1.003 ($\sim$ 1/cos$\phi_0$), which is nearly independent of the chain length $N$~\cite{supple}. Similarly, the differences in calculated specific heat $C_\mathrm{XXZ}^\perp$ and $C_\mathrm{DM}^\perp$ are also completely negligible, compared to the experimental standard errors. Therefore, below, we utilize $\widetilde{\mathcal{H}}$ (Equation (\ref{eq3})) plus $\mathcal{H}_\mathrm{Z}^\parallel$ or $\mathcal{H}_\mathrm{Z}^\perp$ to simulate the magnetization and specific heat down to low temperatures, using the integrable Bethe ansatz method (only available for $H$ $\parallel$ $z$ and $N$ $\to$ $\infty$)~\cite{JPSJ.54.2808,PhysRevLett.54.2131,takahashi2001simplification,takahashi2003thermodynamics} and 100-site ($N$ = 100) SSE quantum Monte Carlo algorithm~\cite{PRE.66.046701,PhysRevB.59.R14157}.

\subsection{Perturbations beyond the nearest neighbors}

\begin{figure}
\begin{center}
  \includegraphics[width=8cm,angle=0]{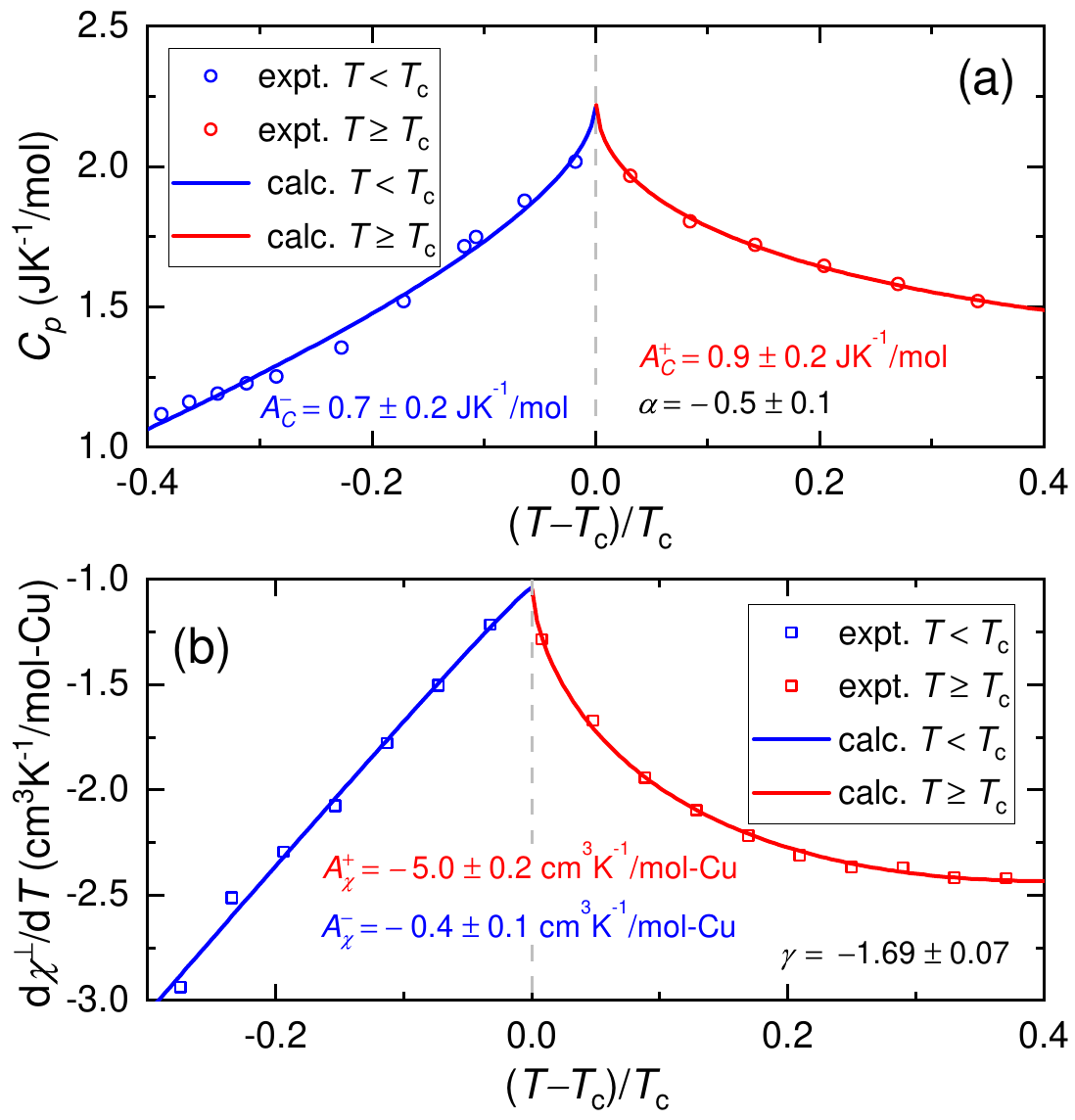}
  \caption{Temperature dependence of (a) zero-field specific heat ($C_p$) and (b) $T$-derivative dc susceptibility (d$\chi^\perp$/d$T$ = d$M^\perp$/d$T$/$H^\perp$, where $\mu_0H^\perp$ = 0.01 T is applied along the $b$ axis) measured on LuCu(OH)$_3$SO$_4$. The specific heat data is fitted to $C_p=\frac{A_C^\pm}{\alpha} \lvert t \rvert^{-\alpha}+B_C+E_C t $ whereas the dc susceptibility data is fitted to d$\chi^\perp$/d$T$ = $\mp\gamma A_\chi^\pm \lvert t \rvert^{-\gamma-1}+B_\chi+E_\chi t$, at $T \geq T_\mathrm{c}$ ($+$) and $T < T_\mathrm{c}$ ($-$), respectively. Here, $A_C^\pm$ and $A_\chi^\pm$ represent critical-behavior coefficients, the fitted critical exponents $\alpha$ and $\gamma$ are listed, $B_C+E_C t$ and $B_\chi+E_\chi t$ are caused by non-critical components in a linear-$t$ approximation, $t\equiv(T-T_\mathrm{c})/T_\mathrm{c}$, and $T_\mathrm{c}$ = 2.5 K.}
  \label{figr2}
\end{center}
\end{figure}

Our semi-quantitative DFT+$U$ simulations~\cite{supple} suggest that the NNN intrachain coupling is weakly antiferromagnetic in LuCu(OH)$_3$SO$_4$, $J_2$/$J_1$ $\sim$ 1\%. The antiferromagnetic $J_2$ induces spin frustration and the notorious sign problem, generally impeding exact resolutions. Fortunately, $J_2$/$J_1$ in LuCu(OH)$_3$SO$_4$ is significantly smaller, $\sim$ 1\%, compared to the critical value $J_2^\mathrm{c}$/$J_1$ = 25\% reported in the theoretical works~\cite{PhysRevB.81.094430,PhysRevB.86.094417}. Furthermore, our many-body simulations confirm the negligible effects of the NNN intrachain coupling ($J_2$ = 1\%$J_1$) on both magnetization and specific heat~\cite{supple}. Hence, we can safely disregard the $J_2$ terms in the effective XXZ Hamiltonian of Eq.~(\ref{eq3}).

Conversely, the minor interchain couplings can trigger a magnetic transition and notably influence observations at low temperatures. Despite the strong ferromagnetic intrachain coupling, the presence of weak but non-negligible interchain couplings induces a three-dimensional (3D) antiferromagnetic order below $T_\mathrm{c}$ = 2.5 K (see below), akin to observations in Li$_2$CuO$_2$~\cite{lorenz2009highly} and Ca$_2$Y$_2$Cu$_5$O$_2$~\cite{PhysRevB.100.104415}. For simplicity, we address these small interchain couplings at a mean-field level for $H^\perp > H_\mathrm{c}^\perp$ [see Fig.~\ref{fig3}(a)],
\begin{multline}
\widetilde{\mathcal{H}^\perp} = -\sqrt{J_1^2+D^2}\sum_j(S_j^xS_{j+1}^x+S_j^yS_{j+1}^y+\Delta S_j^zS_{j+1}^z)
\\-(\mu_0H^\perp g^\perp-6J_\mathrm{p}\langle S^x\rangle)\sum_jS_j^x,
\label{eq4}
\end{multline}
where $J_\mathrm{p}$ = $J_\mathrm{1p}$/3+2$J_\mathrm{2p}$/3. After a sufficient number of self-consistent iterations ($\geq$ 200), we obtain the transverse magnetization $M_\mathrm{calc}^\perp$ = $g^\perp\langle S^x\rangle$ = $g^\perp\langle\Sigma_jS_j^x\rangle$/$N$, which is independent of the iteration number. Combined fits to the transverse magnetization ($M^\perp$/$H^\perp$) measured at $\mu_0H^\perp$ = 1 and 3 T result in the least standard deviation $\sigma$ = $\sqrt{\Sigma_j(M_{\mathrm{expt},j}^\perp/H_j^{\perp}-M_{\mathrm{calc},j}^\perp/H_j^{\perp})^2/N_\mathrm{d}}$ = 0.0358 cm$^3$/mol-Cu at $J_\mathrm{p}$/$J_1$ = 0.21\% [see Fig.~\ref{fig2}(d)]. Here, $N_\mathrm{d}$ represents the number of experimental data points.

Treating $J_\mathrm{p1}$ and $J_\mathrm{p2}$ at the mean-field level reveals two interchain antiferromagnetic orders at low temperatures and magnetic fields: stripe-\uppercase\expandafter{\romannumeral1} order at $J_\mathrm{p1}$ $>$ $J_\mathrm{p2}$ and stripe-\uppercase\expandafter{\romannumeral2} order at $J_\mathrm{p1}$ $<$ $J_\mathrm{p2}$ (see Fig.~\ref{fig3}). Fitting to the critical magnetic field measured at 1.8 K, $\mu_0H_\mathrm{c}^\perp$ = 0.33(2) T, yields two optimal sets of interchain couplings at $J_\mathrm{p}$/$J_1$ = 0.21\%: \uppercase\expandafter{\romannumeral1}. $J_\mathrm{p1}$/$J_1$ = 1.01\% and $J_\mathrm{p2}$/$J_1$ = $-$0.18\%. \uppercase\expandafter{\romannumeral2}. $J_\mathrm{p1}$/$J_1$ = $-$0.21\% and $J_\mathrm{p2}$/$J_1$ = 0.42\% [Fig.~\ref{fig3}(a)]. Both parameter sets agree with the critical temperature $T_\mathrm{c}$ measured at $\sim$ 0 T [see Figs.~\ref{fig3}(b) and~\ref{fig3}(c)] and well reproduce the magnetization/susceptibility measured on the single-crystal sample at 4.2 K [see Figs.~\ref{fig4}(a) and \ref{fig4}(b)]. While we cannot presently distinguish which set of interchain couplings is more applicable for LuCu(OH)$_3$SO$_4$ based solely on the experimental data, the DFT+$U$ calculations suggest the dominant antiferromagnetic nature of $J_\mathrm{p1}$, favoring the parameter set \uppercase\expandafter{\romannumeral1} and thus the stripe-\uppercase\expandafter{\romannumeral1} interchain magnetic order.

\begin{figure*}
\begin{center}
  \includegraphics[width=16cm,angle=0]{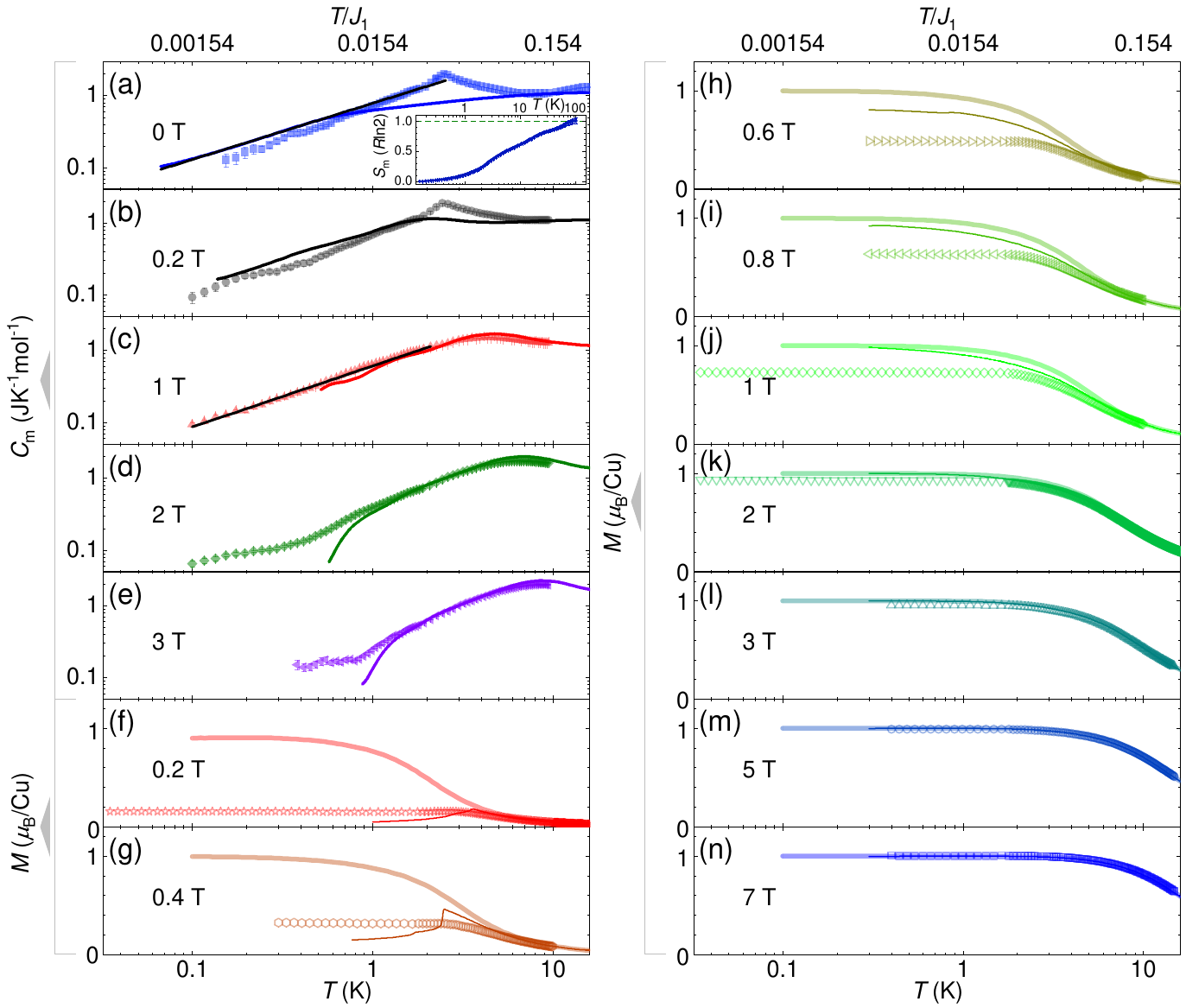}
  \caption{(a)-(e) Low-$T$ magnetic specific heat ($C_\mathrm{m}$) measured on LuCu(OH)$_3$SO$_4$ powder. The lattice contribution is carefully evaluated and subtracted from the total measured specific heat~\cite{supple}. The zero-field specific heat is calculated using the Bethe ansatz method [blue line, see (a)], while the nonzero-field data are computed by the SSE method [colored lines, see (b)-(e)]. Here, $J_1$ = 65 K and $\Delta$ = 0.994 (ideal XXZ-chain model) are used, and the nonzero-field calculations are powder-averaged, $C_\mathrm{m,calc}$ = 2$C_\mathrm{m,SSE}^\perp$/3+$C_\mathrm{m,Bethe}^\parallel$/3~\cite{supple}. The black line shows a power-law fit to the experimental data measured at 1 T (red triangles) below 2 K in (c), $C_\mathrm{m}$ $\propto$ $T^{0.848(7)}$, whereas the black line in (a) displays a power-law fit to the zero-field calculated (Bethe) data between 0.1 and 0.6 K, $C_\mathrm{m,Bethe}$(0 T) $\propto$ $T^{0.782(2)}$. The inset of (a) displays the zero-field magnetic entropy integrated from the experimental data, $S_\mathrm{m}(T)$ = $\int_{\mathrm{0.15 K}}^TC_\mathrm{m}(T')dT'/T'$. (f)-(n) Low-$T$ magnetization measured on LuCu(OH)$_3$SO$_4$ powder. The thick and thin lines respectively present SSE calculations using the ideal XXZ-chain model ($J_1$ = 65 K and $\Delta$ = 0.994) and detailed model (with additional $J_\mathrm{p1}$/$J_1$ = 1.01\% and $J_\mathrm{p2}$/$J_1$ = $-$0.18\%) in a powder average. In the detailed model, interchain interactions are treated by a self-consistent mean-field approximation.}
  \label{fig5}
\end{center}
\end{figure*}

Our self-consistent mean-field SSE method accurately simulates the experimental data at $T$ $\geq$ 4.2 K ($>$ $T_\mathrm{c}$) or $\mu_0H^\perp$ $\geq$ 1 T ($>$ $\mu_0H_\mathrm{c}^\perp$), where the symmetry of the spin system is fully recovered by thermal fluctuations or Zeeman interaction [see Figs.~\ref{fig2}(d),~\ref{fig3},~\ref{fig4}(a), and \ref{fig4}(b)]. Moreover, the mean-field model also shows excellent agreement with the boundary of $\mu_0H_\mathrm{c}^\perp$($T$) and $T_\mathrm{c}$($H^\perp$) in the $B$-$T$ phase measured on LuCu(OH)$_3$SO$_4$~\cite{supple}. Nevertheless, the calculated magnetic-field-induced transition is considerably sharper than the experimental observations at 1.8 K [see Fig.~\ref{fig4}(b)]. Additionally, the mean-field calculations fall short in accurately reproducing the experimental data at low temperatures and magnetic fields (within or near the interchain ordered phase). These discrepancies may stem from neglecting intrachain quantum spin fluctuations in the mean-field treatment of interchain couplings.

In LuCu(OH)$_3$SO$_4$, the putative interchain antiferromagnetic transition is likely significantly suppressed by the strong quantum spin fluctuations along the chain, as indicated by the weak kinks in susceptibility [d$M^\perp$/d$H^\perp$, Fig.~\ref{fig4}(b)], magnetization [Figs.~\ref{fig4}(c) and~\ref{fig4}(d)], and specific heat [Fig.~\ref{figr2}(a)] at critical points. The observed peaks in d$M^\perp$/d$H^\perp$ and d$M^\perp$/d$T$/$H^\perp$ [Figs.~\ref{fig4}(b) and \ref{fig4}(d)] suggest that these temperature- and field-induced magnetic transitions are of second-order. The absence of frequency dependence in ac susceptibilities, the lack of splitting between ZFC and FC dc susceptibilities [Fig.~\ref{fig4}(c)], and the absence of evident magnetic hysteresis in magnetization [Fig.~\ref{fig4}(a)] exclude the possibilities of glassy or ferromagnetic nature for these transitions. Finally, as shown in the $B$-$T$ phase diagram~\cite{supple}, a weak applied field above $\sim$ 0.6 T, corresponding to an energy of $\mu_0Hg\mu_\mathrm{B}$ $\sim$ 1\%$J_1$, is large enough to overcome the interchain antiferromagnetic correlations driven by $J_\mathrm{p1}$/$J_1$ = 1.01\% and $J_\mathrm{p2}$/$J_1$ = $-$0.18\% (as discussed above).

Following Refs.~\cite{kornblit1973heat,PhysRevB.11.2678,wosnitza2007thermodynamically}, we conducted traditional Kornblit-Ahlers-Buehler fits to the experimental thermodynamic properties of LuCu(OH)$_3$SO$_4$ near $T_\mathrm{c}$ = 2.5 K, as depicted in Fig.~\ref{figr2}. However, the fitted critical exponents, $\alpha$ $\sim$ $-$0.5 and $\gamma$ $\sim$ $-$1.7, are significantly smaller than the theoretically predicted values for 3D systems~\cite{wosnitza2007thermodynamically}. This discrepancy suggests much broader peaks observed in both $C_p$ and d$\chi$/d$T$ (or $\chi$) of LuCu(OH)$_3$SO$_4$ at $T_\mathrm{c}$ = 2.5 K, indicating very unconventional critical behaviors, likely influenced by strong intrachain spin fluctuations. We emphasize that the negative sign of $\gamma$ is a special novel feature of LuCu(OH)$_3$SO$_4$, compared to other compounds~\cite{wosnitza2007thermodynamically}. This is confirmed by the measured raw susceptibilities, which are (nearly) monotonic functions of temperature, and exhibit only a weak kink at $T_\mathrm{c}$ = 2.5 K (similar to the specific heat, see Fig.~\ref{fig5}(a)), as shown in Fig.~\ref{fig4}(c). The measured values of both $\alpha$ and $\gamma$ can determine the type of the phase transition, which may be useful in future theoretical and experimental investigations.

The weak but nonzero interchain antiferromagnetic couplings could induce a 3D N\'{e}el order (see Fig.~\ref{fig3}). LuCu(OH)$_3$SO$_4$ only approximates the pure XXZ-chain model, and nonzero interchain couplings are inevitable. As illustrated in Fig.~\ref{fig4}(d), $T^*$ $\sim$ 3.4 K denotes the higher temperature at which the $T$-derivative magnetization/susceptibility d$M$/d$T$/$H$ exhibits anomalies. It approaches the critical temperature $T_\mathrm{c}$ = 2.5 K, prompting us to define it as the onset temperature of the proposed 3D N\'{e}el magnetic order (see Fig.~\ref{fig3}).

\subsection{Proximate Tomonaga-Luttinger liquid state}

The above observations and simulations suggest that the perturbation interactions beyond the nearest neighbors are weak and only have a marginal effect on the quantum magnetism of LuCu(OH)$_3$SO$_4$. Since these perturbations are inherent and completely unavoidable in any real materials, we propose LuCu(OH)$_3$SO$_4$ as the most perfect candidate so far for the ideal spin-1/2 ferromagnetic XXZ chain model [see Eq.~(\ref{eq3}) with $\mathcal{H}'$ $\sim$ 0]. Therefore, below we focus on the direct comparisons between the low-$T$ observations and theoretical expectations of the ideal exactly solvable model (Fig.~\ref{fig5}). When the chain length $N$ is less than $\sim$ 30, the ED/SSE simulations exhibit evident finite-size effects below $\sim$ 8 K ($\sim$ 0.1$J_1$) despite the periodic boundary condition employed~\cite{supple}, suggesting highly exotic ground-state properties of the spin-1/2 ferromagnetic XXZ chain system with long-distance unconventional entanglements.

It remains technically difficult to measure the single-crystal specific heat and mK magnetization, limited by the crystal size~\cite{supple}. In the following discussions, we rely on the low-$T$ data measured on the ground powder. LuCu(OH)$_3$SO$_4$  is suggested to approximate a gapless Tomonaga-Luttinger liquid ground state of the ideal spin-1/2 ferromagnetic XXZ chain model, supported by: 1. Both our experimental data (Figs.~\ref{fig2} and \ref{fig3}) and DFT+$U$ calculations consistently demonstrate $\Delta$ $\sim$ 0.994 and $J_2$/$J_1$ $\sim$ 1\% ($\ll$ 1/4), suggesting LuCu(OH)$_3$SO$_4$ is deeply inside the Tomonaga-Luttinger-liquid-ground-state region~\cite{PhysRevB.81.094430,PhysRevB.86.094417}. 2. The magnetic entropy measured at 0 T approaches zero at low temperatures ($<$ 2 K) from the full value of $R$ln(1+2$S$) = $R$ln2 at high temperatures ($\sim$ 100 K), where $S$ = 1/2 [see inset of Fig.~\ref{fig5}(a)]. This indicates the spin-1/2 (Cu$^{2+}$) system of LuCu(OH)$_3$SO$_4$ indeed approaches its ground-state properties below $\sim$ 2 K ($\sim$ 3\%$J_1$). 3. Below $\sim$ 2 K, the magnetic specific heat exhibits a power-law temperature dependence down to the lowest measured temperature [$\sim$ 0.1 K, see Fig.~\ref{fig5}(c)], $C_m$ $\propto$ $T^{0.85}$, at low applied fields of $\leq$ 1 T ($\mu_0Hg\mu_\mathrm{B}$ $\leq$ 2\%$J_1$). Moreover, the dc magnetization measured at $\mu_0H$ = 0.2 T flattens out to an unsaturated value ($<$ $g$/2) below $\sim$ 2 K. These observations strongly suggest the gapless nature of the low-lying spin excitations in LuCu(OH)$_3$SO$_4$. 4. The ideal spin-1/2 ferromagnetic XXZ chain model ($J_1$ = 65 K and $\Delta$ = 0.994) yields the exactly solved specific heat between $\sim$ 0.1 and 0.6 K, $C_\mathrm{m,Bethe}$(0 T) $\propto$ $T^{0.782(2)}$ [Fig.~\ref{fig5}(a)], in good agreement with the experimental findings. The calculated magnetization at 0.2 T also flattens out at low temperatures, qualitatively consistent with the experimental result [Fig.~\ref{fig5}(f)].

The applied magnetic field can shift the nuclear Schottky specific heat to higher temperatures~\cite{PhysRevB.97.184434}, possibly contributing to the larger distinctions between the experimental and calculated data at low temperatures in 2 and 3 T [Figs.~\ref{fig5}(d) and \ref{fig5}(e)]. Interchain interactions have a minor effect on specific heat, except for the weak peak centered at $T_\mathrm{c}$ $\sim$ 2.5 K observed at $\mu_0H$ $<$ 1 T. In contrast, interchain interactions ($J_\mathrm{p1}$ and $J_\mathrm{p2}$) significantly suppress magnetization at low temperatures and low applied fields, providing a better explanation for the experimental findings [Figs.~\ref{fig5}(f)-\ref{fig5}(n)]. Since interchain interactions are treated only at the mean-field level, and a powder average is adopted to lower the computational cost, achieving full quantitative agreement between experimental and simulated magnetization data at low temperatures and applied fields remains extremely challenging. Improved treatment of interchain interactions and larger-size single crystals of LuCu(OH)$_3$SO$_4$ are highly required for further studies.

\section{Conclusions}

We propose LuCu(OH)$_3$SO$_4$, a structurally disorder-free compound, as a promising candidate for the exactly solvable spin-1/2 ferromagnetic XXZ chain model ($J_1$ = 65 K and $\Delta$ = 0.994). Determined small values for the interchain and NNN intrachain interactions ($J_\mathrm{p1}$/$J_1$ $\sim$ 1.01\%, and $J_\mathrm{p2}$/$J_1$ $\sim$ $-$0.18\%, and $J_2$/$J_1$ $\sim$ 1\%) support this candidacy. Measurements of specific heat and magnetization down to $T$ = 35 mK ($\sim$ 0.0005$J_1$) exhibit gapless ground-state behaviors, aligning with theoretical expectations for the ideal model. These observations indicate that LuCu(OH)$_3$SO$_4$ approximates a gapless Tomonaga-Luttinger liquid state at low temperatures. Our findings mark an initial step towards understanding the potential Tomonaga-Luttinger liquid phase in LuCu(OH)$_3$SO$_4$ and offer new insights for experimental investigations into exactly solvable strongly correlated quantum systems.

\bigbreak

\begin{acknowledgments}
We gratefully acknowledge Haijun Liao for helpful discussion. This work was supported by the National Key R\&D Program of China (Grant No. 2023YFA1406500), the National Natural Science Foundation of China (No. 12274153), and the Fundamental Research Funds for the Central Universities (No. HUST: 2020kfyXJJS054).
\end{acknowledgments}

\bigbreak

\bibliography{FMchain}

\clearpage

\addtolength{\oddsidemargin}{-0.75in}
\addtolength{\evensidemargin}{-0.75in}
\addtolength{\topmargin}{-0.725in}

\newcommand{\addpage}[1] {
 \begin{figure*}
   \includegraphics[width=8.5in,page=#1]{Supplementary_R2.pdf}
 \end{figure*}
}
\addpage{1}
\addpage{2}
\addpage{3}
\addpage{4}
\addpage{5}
\addpage{6}
\addpage{7}
\addpage{8}

\end{document}